\documentclass[12pt,twoside]{article}
\newcommand{\bls}[1]{\advance\baselineskip -#1pt}

\newcommand{\starting}{\topmargin=-2.5cm
\renewcommand{\@evenfoot}{\roman{page} \hfil}
\renewcommand{\@oddfoot}{\hfil \roman{page}}}
\newdimen{\Ione}   \newdimen{\Ipoltora}  \Ione=13pt  \Ipoltora=15pt
\def\Title#1#2#3{%
\baselineskip=\Ipoltora
\begin{center}  {\Large\bf  \uppercase{#1} \\ }
                 \bigskip\bigskip  {#2} \\  {#3} \\ \end{center}}
\long\def\Abstract#1{%
\bigskip
\parbox{0.93\textwidth}{%
\begin{center}   {\bf Abstract} \\   \end{center}

\medskip    {    \baselineskip=\Ione

#1

}    \vss   }   \bigskip    }
\def\@cite#1#2{{$^{\hbox{\scriptsize#1})}$\if@tempswa , (see #2)\fi}}
\def\thebibliography#1{\section*{{\uppercase{\refname}}}\list
 {\arabic{enumi}}{\settowidth\labelwidth{[#1]}\leftmargin\labelwidth
 \advance\leftmargin\labelsep
 \usecounter{enumi}}
 \def\newblock{\hskip .11em plus .33em minus .07em}
 \sloppy\clubpenalty4000\widowpenalty4000
 \sfcode`\.=1000\relax}

\begin{document}

    \setcounter{equation}{0}

\Title{Quantum field theory in curved space-time and the early Universe}
{A.A. Grib$^1$ ,
Yu.V. Pavlov$^2$ }
{$^1$A.Friedmann Laboratory for Theoretical Physics,  \\
30/32 Griboedov can, St.Petersburg, 191023, Russia,  \\
   e-mail:  grib@friedman.usr.lgu.spb.su      \\
$^2$Institute of Mechanical Engineering, Russian Academy of Sciences, \\
   61 Bolshoy, V.O., St.Petersburg, 199178, Russia,    \\
   e-mail:  pavlov@ipme.ru}

%

\Abstract{%
    New results on finite density of particle creation for nonconformal
massive scalar particles in Friedmann Universe as well as new counterterms
in dimensions higher than 5 are presented.
    Possible role of creation of superheavy particles for explaining
observable density of visible and dark matter is discussed.  }


\section{Introduction}
%
Quantum field theory in curved space-time is the generalization of well
developed theory in Minkowski space-time, so it has some features of
the standard theory together with new ones due to the curvature.
  It was actively developed in the 70-ties of the last century
(see our books~\cite{GMM1,GMM}) however some new results were obtained
just recently and in this paper we'll concentrate on these new results
as well as on some physical interpretation of the old ones.

   One of the main results of the theory for the early Friedmann Universe
was the calculation of finite stress-energy tensor for particle creation.
Let us begin from some remarks on what is meant by particle creation in
curved space-time?
    In spite of the absence of the standard definition of the particle
as the representation of the Poincare group in curved space-time
where Poincare group is not a group of motions, one still has physically
clear idea of the particle.
The particle is understood as the classical point like object moving along
the geodesic of the curved space-time.
This classical particle however is the quasiclassical approximation of some
quantum object, so the main mathematical problem is to find the answer
to the question: what is the Fock quantization of the field giving just
this answer for particles in this approximation?
The answer to this question was given by us in the 70-ties for conformal
massive scalar particles and spinor particles in Friedmann space-time
by use of the metrical Hamiltonian diagonalization method.
Creation of particles in the early Universe from vacuum by the gravitational
field means that for some small time close to singularity the stress-energy
tensor has the geometrical form, expressed through some combinations of
the Riemann tensor and its derivatives while for the time larger than the
Compton one it has the form of the dust of pointlike particles with mass
and spin defined by the corresponding Poincare group representation.

    So our use of quantum field theory in curved space-time with its methods
of regularization (dimensional regularization and
Zeldovich-Starobinsky regularization) was totally justified by
the obtained results. However still unclear was the situation with
the minimally coupled scalar field where our method led to
infinite results for the density of created particles, it is only
recently that Yu.V.Pavlov~\cite{Pv} could find the transformed
Hamiltonian, diagonalization of which leads to finite results for
this case. Also for higher dimensions it was found~\cite{Pv2} that
new counterterms different from those in 4-dimensional space-time
appear in the Lagrangian.

\section{Some remarks on the physical interpretation of terms in vacuum
polarization dependent on mass}
%
    Our calculations of the vacuum expectation value of the stress-energy
tensor of the quantized conformal massive scalar, spinor and vector
fields~\cite{GMM1,GMM} led to the expressions different for strong and weak
external gravitational field.
   By strong field one understands the field with the curvature much larger
than the one defined by the mass of the particle.
For strong gravitational field it is basically polarization of vacuum terms
which becomes zero if the gravitation is zero, for weak gravitation it is
defined by particles created previously, so that if gravitation becomes
zero particles still exist.
Vacuum polarization for strong gravitation consists of three terms:
the first is due to the conformal anomaly and it does not depend on mass
of the particle at all, the second is due to the Casimir effect and is
present for the closed Friedmann space, the third is vacuum polarization
dependent on mass of the particle which is for scalar conformal particles
    \begin{equation}
<\!T_{ik}\!>_m=\frac{m^2}{144 \pi^2}\,G_{ik} +
\frac{m^4}{64 \pi^2}\,g_{ik} \ln \biggl(\frac{\Re}{m^4}\biggr)\,.
\label{Tikm}
    \end{equation}
Here $G_{ik}$ is the Einstein tensor, $g_{ik}$ is the metrical
tensor, $\Re$ is some geometrical term of the dimension of the
fourth degree of mass. It is interesting that the first term is
the illustration of the Sakharov's idea of gravitation as the
vacuum polarization. If one takes the Planckean mass one has just
the standard expression for the term in the Einstein equation. One
can also think that gravitation is the manifestation of the vacuum
polarization of all existing fields with different masses. However
for weak gravitation for the time of evolution of the Friedmann
Universe large than the Compton one defined by the mass there is
no Sakharov term, it is compensated. It is just manifestation of
the fact, known also in quantum electrodynamics that vacuum
polarization is different in strong and weak external fields being
asymptotics of some general expression. So the physical meaning of
this term in strong field is finite change of the gravitational
constant, so that the new effective gravitational constant is
different from that in the weak field, being
    \begin{equation}
\frac{1}{8 \pi G_{eff} } = \frac{1}{8 \pi G} +
\frac{m^2}{144 \pi^2} \,.
    \end{equation}

    The second term on the right side~(\ref{Tikm}) describes what is now
called ``quintessence" -- cosmological constant, dependent on time.
   This ``quintessence'' is different for different stages of
the evolution of the Friedmann Universe.
    We calculated it previously for the radiation
dominated Universe, so that together with the Sakharov term one has
    \begin{equation}
<\!T_{00}\!>_m=-\frac{m^2}{48 \pi^2} \biggl( \frac{a'}{a^2} \biggr)^2 -
\frac{m^4}{16 \pi^2} \left\{ \ln(ma) + C + \frac{1}{4} +
\frac{1}{a^4} \int \limits_0^\eta \!d\eta_1 \frac{d a^2}{d\eta_1}
\int \limits_0^\eta \! d\eta_2 \frac{d a^2}{d\eta_2} \ln|\eta_1\!-\eta_2|
\right\}.
    \end{equation}
Here $a$ is the scale factor of the Friedmann space-time, $\eta $ is the
conformal time, $C=0.577 \ldots $ is the Euler constant.

\section{The minimally coupled scalar field in homogeneous isotropic space}
%
    We consider a complex scalar field with Lagrangian
\begin{equation}
L(x)=\sqrt{|g|}\ [\,g^{ik}\partial_i\varphi^*\partial_k\varphi -(m^2+\xi R)
\,\varphi^* \varphi \,] \,,
\label{Lag}
     \end{equation}
where $g={\rm det} \{ g_{ik} \}$, $R$ is the scalar curvature, $\xi$ is
a coupling to the curvature, which is equal to zero for the case of
minimal coupling.
    The metric of $N$-dimensional  homogeneous isotropic space-time is
\begin{equation}
 ds^2= a^2(\eta)\, (d\eta^2 - \gamma_{\alpha \beta} dx^\alpha dx^\beta) \,.
 \label{ds}
     \end{equation}
Here $\gamma_{\alpha \beta}$ is the metric of $(N-1)$-dimensional space of
constant curvature $K=0, \pm 1$\,.

     Let us take a new Lagrangian different from the previous by
the $N$-divergence
$ L^{\Delta}(x)=L(x)+ \partial J^i / \partial x^i $,
where $(J^i)$ in coordinate system $(\eta, {\bf x})$ is given by
     \begin{equation}
(J^i)=(\sqrt{\gamma}\,c\,\tilde{\varphi}^*\, \tilde{\varphi}\,
(N\!-\!2)/2,\, 0, \, \ldots \,, 0),  \ \ \ \ \
 \tilde{\varphi}(x)= a^{(N-2)/\,2}(\eta) \, \varphi(x)  \,.
\label{02}
    \end{equation}
Here $c=a'/a, \ \ \gamma={\rm det}\{\gamma_{\alpha \beta} \} $.
   Lagrangian  $ L^{\Delta}(x) $ leads to the canonical Hamiltonian for
the transformed field $\tilde{\varphi}$
        \begin{equation}
H(\eta) = \int d^{N-1}x\,\sqrt{\gamma} \, \biggl\{
\tilde{\varphi}^{* \prime} \tilde{\varphi}'
+\gamma^{\alpha \beta}\partial_\alpha\tilde{\varphi}^*
\partial_\beta\tilde{\varphi}
+ \biggl[m^2a^2-\Delta \xi\, a^2 R +
\biggl( \frac{N-2}{2}\biggr)^2 K \biggr]\,
 \tilde{\varphi}^* \tilde{\varphi} \biggr\} ,
\label{Hm}
\end{equation}
    where $\Delta \xi =(N-2)/[4(N-1)] - \xi $.

    For quantizing, we expand $\tilde{\varphi}(x)$  with respect to
the complete set of motion equation solutions

            \begin{equation}
\tilde{\varphi}(x)=\int \! d\mu(J)\,\biggl[\,\tilde{\varphi}^{(-)}_J
\,a^{(-)}_J + \tilde{\varphi}^{(+)}_{J}\,a^{(+)}_{J}\,\biggr],   \ \ \
\tilde{\varphi}^{(+)}_J (x)=
\frac{g_\lambda(\eta)\,\Phi^*_J({\bf x})}{\sqrt{2}} ,      \ \ \
\tilde{\varphi}^{(-)}_J (x)=\tilde{\varphi}^{(+)*}_J (x)  ,
\label{11}
\end{equation}
where  $d \mu(J)$ is a measure in the space of eigenvalues of the
Laplace-Beltrami operator $\Delta_{N\!-\!1}$ on
$(N-1)$-dimensional space $\{x^\alpha\}$,
    \begin{equation}
 g_\lambda''(\eta) + \Omega^2(\eta) g_\lambda(\eta) =0, \ \ \
\Omega^2(\eta)=m^2 a^2 + \lambda^2 -\Delta \xi\,a^2 R  , \ \ \
\Delta_{N\!-\!1} \Phi_J = -\biggl(\lambda^2 -
\frac{(N\!-\!2)^2}{4} K \biggr) \Phi_J.
\label{10}
\end{equation}

    The initial conditions  corresponding to
the diagonal form of Hamiltonian~(\ref{Hm}) in operators
$\stackrel{*}{a}\!{}_J^{(\pm)}$ and $ a_J^{(\pm)} $ in time  $\eta_0 $ are
    \begin{equation}
g_\lambda'(\eta_0)=i\, \Omega(\eta_0)\, g_\lambda(\eta_0) \,, \ \ \
|g_\lambda(\eta_0)|= 1/\sqrt{\Omega(\eta_0) } \ .
\label{Ic}
\end{equation}
   The density  of the created pairs of particles~\cite{GMM}
(for $N=4$) is
     \begin{equation}
n(\eta)=\frac{1}{2 \pi^2 a^3(\eta)}
\int d\lambda\,\lambda^2 \,S_\lambda(\eta) \,,
\label{33}
\end{equation}
     where
     \begin{equation}
S_\lambda(\eta)=
\frac{1}{4\Omega}\biggl( |g_\lambda|^2+\Omega^2 |g_\lambda|^2 \biggr) -
\frac{1}{2} \,.
\label{be2}
\end{equation}
    It can be shown~\cite{Pv}, that
$ S_\lambda(\eta) \sim \lambda^{-6} ,\ \lambda \to \infty, \ \
\forall \, \xi $.
   That is why the density of created particles of scalar field
with arbitrary $\xi$ in homogeneous isotropic 4-dimensional space-time
is finite.
    So it is shown differently from~\cite{BD}, that even for minimal
coupling the result is finite!

\section{The renormalization for scalar field in $N$-dimensional
quasi-Euclidean space-time}
%
    The vacuum expectation value of the stress-energy tensor of scalar field
with Lagrangian~(\ref{Lag}) for vacuum corresponding to
conditions~(\ref{Ic}) on $N$-dimensional space-time with
$\gamma_{\alpha \beta}=\delta_{\alpha \beta}$ in~(\ref{ds}) are
     \begin{equation}
<\!0\,|\,T_{ik}|\,0\!>=\frac{B_N}{a^{N-2}} \int \limits_{0}^{\infty}
\! d\lambda \,\lambda^{N-2} \tau_{ik}  \,,
\label{Ttik}
\end{equation}
    where
$ B_N\!=\!\left[ 2^{N-3}\pi^{(N-1)/\,2} \, \Gamma((N\!-\!1)/2) \right]^{-1}$,
$ \ \ \Gamma(z)$ is the gamma-function,
             \begin{equation}
\tau_{00}=\Omega \biggl( S+\frac{1}{2}\biggr) +
\Delta \xi \,(N\!-\! 1) \left[ c \, V + \biggl( c'+(N\!-\! 2) \, c^2\biggr)
\frac{1}{\Omega} \left(\!S+\frac{1}{2}  U +
\frac{1}{2}\right)\right] ,
\label{ts00}
\end{equation}
                      \begin{eqnarray}
\tau_{\alpha \beta} &=& \delta_{\alpha \beta}
\left\{\frac{\lambda^2}{(N-1)\,\Omega} \left( S+\frac{1}{2} \right)-
\frac{\Omega^2-\lambda^2}{2 \,(N-1)\,\Omega} \, U - \right.
\nonumber   \\
&-& \Delta \xi \left[ (N\!-\!1)\, \frac{c'}{\Omega}
\left. \left( S+\frac{1}{2}\, U +\frac{1}{2} \right) +
2\,\Omega \,U\! -\! (N\!-\!1)\,c\,V \right] \right\} ,
\label{tsab}
\end{eqnarray}
    \begin{equation}
S=\frac{|g_\lambda'|^2 + \Omega^2\,|g_\lambda|^2}{4 \, \Omega}
-\frac{1}{2} \ , \ \ \ \
U=\frac{\Omega^2 \, |g_\lambda|^2- |g_\lambda'|^2}{2\, \Omega} \, \ \ \ \
V= - \frac{d (g_\lambda^* g_\lambda)}{2\, d \eta} \,.
\label{SUV}
\end{equation}

   The vacuum expectation~(\ref{Ttik}) has
$[N/2]+1 $ different types of divergences
($[b]$ denotes the integer part of $b$):
$\ \sim \lambda^N, \ \lambda^{N-2}, \ldots , \ln \lambda $ \ \
if $N$ is even, and
$ \sim \lambda^N, \ \lambda^{N-2}, \ldots , \lambda $ \ \
if $N$ is odd.
   For renormalization we are using generalization for $N$-dimensional
case~\cite{Pv2} of $n$-wave procedure~\cite{ZlSt}
    \begin{equation}
<\!0\,|\,T_{ik}|\,0\!>_{\! ren}=\frac{B_N}{a^{N-2}}
\int \limits_{0}^{\infty} \! d\lambda \,\lambda^{N-2}
\biggl[\, \tau_{ik} -\sum \limits_{l=0}^{[N/2]} \tau_{ik}[l]\,\biggr] \,,
\label{Trik}
\end{equation}
        \begin{equation}
\tau_{ik}[l] =\frac{1}{l!} \lim_{n \to \infty }
\frac{\partial^{\, l}}{\partial (n^{-2})^l}
\left( \frac{1}{n} \, \tau_{ik} (n \lambda, n m) \right).
\label{taul}
\end{equation}

   The geometrical structure of counterterms has been determined with
help of the dimensional regularization. It can be shown~\cite{Pv2}, that
 in the case $N=4,5$ all counterterms correspond to renormalization
 of constants in the bare gravitational Lagrangian of form
    \begin{equation}
L_{gr,\,0}=\sqrt{|g|} \left[ \frac{1}{16 \pi G_0} \,
( R\! -\!2 \Lambda_0 ) +
\alpha_0 \, \biggl( R^{ik} R_{ik} - \frac{N \, R^2}{4(N\!-\!1)}  \, \biggr) +
\beta_0 \, R^2 \right],
\label{Lgr0}
\end{equation}
    where $R_{ik}$ is Ricci tensor.
   Constant $\alpha_0$ has infinite renormalization
under $N=4-\varepsilon, \ \varepsilon \to 0$.

    New counterterms appear for $N \ge 6$ in comparison with a cases $N=4,5$,
for example
     \begin{equation}
\tau_{00}[3]=\omega \, S_6, \ \ \ \
\tau_{\alpha \beta}[3]=\frac{\delta_{\alpha \beta}}{(N-1)\, \omega}
\Biggl[ \lambda^2 \, S_6 - \frac{m^2 a^2}{2} \, U_6 \Biggr] \,,
\label{t0a3}
\end{equation}
   where it's suggested that $\Delta \xi =0 $,
           \begin{equation}
S_6=\frac{5}{32}\, W^6 - \frac{5}{8}\, W^2 (D W)^2 -
\frac{5}{4}\, W^3 D^2 W - \frac{1}{2} D W D^3 W
+ \frac{1}{2} W D^4 W +\frac{1}{4} \biggl( D^2 W \biggr)^2 \,,
\label{S6}
\end{equation}
           \begin{equation}
U_6=\frac{15}{8} \, W^4 DW -  \frac{5}{2}\,(D W)^3 - 10 \, W DW D^2 W -
\frac{5}{2}\, W^2 D^3 W + D^5 W \,,
\label{U6}
\end{equation}
      $ \omega=(m^2a^2+\lambda^2)^{1/2} \,, \ \
W=\omega'/(2 \omega^2) \,, \ \ D=(2 \omega)^{-1} (d\,.../d\eta)$.
   For $N=6,7$ from dimensional analysis of counterterms we can suppose
that
\begin{equation}
\Delta L_{gr,\,0}=\sqrt{|g|} \left[ \gamma_0 R^3 +
\nu_0 R R_i^{\, k} \! R_k^{\, i} +
\zeta_0 R_i^{\, l} R_{\, l}^{\, k} R_k^{\, i} +
\rho_0 \nabla^{\, l}\! R \, \nabla_{\! l} R  + \ldots \right] .
\end{equation}

\section{Superheavy particles in Friedmann cosmology and the dark matter
problem}
%
    It is well known~\cite{GMM,GD} that creation of superheavy particles
with the mass of the order of Grand Unification
$M_X \approx 10^{14} - 10^{15} $\, Gev with consequent decay on quarks and
leptons with baryon charge and CP -- nonconservation is sufficient for
explanation of the observable baryonic charge of the Universe.
   Recently in papers \cite{K3,K4} the possibility of explanation of
experimental facts on observation of cosmic ray particles with the energy
higher than the Greizen-Zatsepin-Kuzmin limit was discussed.
    The proposal is to consider the decay of superheavy particles
with the mass of the order $M_X$\,.
   One can even consider the hypothesis that all dark matter consists
of neutral $X $~-- particles with very low density.
    So the problem is in the numerical estimate of the parameters of the
effective Hamiltonian similar to the theory of $K^0$-mesons
leading  to the observable data.
    Short living $ X^0 $~-- bosons decay on quarks and leptons in time
close to singularity, long living $ X^0 $ -- bosons exist today as the
dark matter.     Here we shall give this estimate.

   Particles are created by gravitation at the Compton time
$ t \sim M^{-1} $ and for $ t \gg M^{-1} $ one has nonrelativistic
gas of created particles with the energy density calculated for the
radiation dominated Friedmann model
$a(t)=a_0 \, t^{1/2} $\,:~\cite{GMM}
\begin{equation}
\varepsilon^{(0)}=2b^{(0)}\,M(M/t)^{3/2} \,,
\end{equation}
    where $b^{(0)} \approx 5.3\cdot 10^{-4} $.
Total number of created particles in the Lagrange volume is
\begin{equation}
N=n^{(0)}(t)\,a^3(t)=b^{(0)}\,M^{3/2}\,a_0^3 \ .
\end{equation}
    In spite of the cosmological order of the number of created
$X $ -- particles
($ N \sim 10^{80} $ for $ M_X \sim 10^{15} $\,Gev~\cite{GMM})
their back reaction on the background metric is small.~\cite{GMM}
    However for ${t \gg M^{-1}} $ there is an era of going from the
radiation dominated model to the dust model of superheavy particles for
$ \varepsilon_{bground}~\approx~\varepsilon^{(0)} $\,,
\begin{equation}
t_X\approx \left(\frac{3}{64 \pi \, b^{(0)}}\right)^2\,
\left(\frac{M_{Pl}}{M_X}\right)^4\,\frac{1}{M_X}  \,.
\end{equation}
    If $M_X \sim 10^{14} $\,Gev,  $\ t_X \sim 10^{-15} $\,sec, if
$M_X \sim 10^{13} $\,Gev --- $t_X \sim 10^{-10} $\,sec.
    So the life time of short living $X $ -- mesons must be smaller then
$t_X $\,. It is evident that if all created  $X $ -- particles were stable,
than the closed Friedmann model could quickly collapse, while all other
models are strongly different from the observable Universe.
   Let us define $d $ -- the permitted part of long living
$X $ -- mesons --- from the condition: on the moment of
recombination $t_{rec} $ in the observable Universe one has
    \begin{equation}
d\,\varepsilon_X(t_{rec}) =\varepsilon_{crit}(t_{rec})  \,, \ \ \ \
d=\frac{3}{64 \pi \, b^{(0)}}\left(\frac{M_{Pl}}{M_X}\right)^2\,
\frac{1}{\sqrt{M_X\,t_{rec}}}\, .
\label{d}
\end{equation}
For $M_X=10^{13} - 10^{14} $\,Gev one has
$d \approx 10^{-12} - 10^{-14} $\,.
    Using the estimate for the velocity of change of the concentration of
long living superheavy particles~\cite{BBV}
$|\dot{n}_x| \sim 10^{-42}\, \mbox{cm}^{-3}\,\mbox{sec}^{-1} $,
and taking the life time $\tau_l $ of long living particles as
$2\cdot 10^{22} $\,sec, we obtain concentration
$n_X \approx 2\cdot 10^{-20} \,\mbox{cm}^{-3} $ at the modern epoch,
corresponding to the critical density for $M_X=10^{14} $\,Gev\,.

    Now let us construct the toy model which can give: \ \
a) short living $X $ -- mesons decay in time
   $\tau_q < 10^{-15} $\,sec, (more wishful is
   $\tau_q \sim 10^{-38} - 10^{-35} $\,sec),
   long living mesons decay with $\tau_l > t_U \approx 10^{18}$\,sec
   \  ($t_U $~--~age of the Universe),         \ \
b) one has small $ d \sim 10^{-14} - 10^{-12} $ part of long living
   $X $ -- mesons, forming the dark matter.

   Baryon charge nonconservation with CP -- nonconservation in full analogy
with the $K^0 $ -- meson theory with nonconserved hypercharge and
CP -- nonconservation leads to the effective Hamiltonian of the decaying
$X, \bar{X} $ -- mesons with nonhermitean matrix.

   The matrix of the effective Hamiltonian is in standard
notations
\begin{equation}
H=     \left(
\begin{array}{cc}
H_{11}  & H_{12}   \\
H_{21}  & H_{22} \\
\end{array}        \right) .
\label{H}
\end{equation}
    Let $H_{11}\! =\! H_{22}$
(due to $CPT$-invariance).
Denote
$\ \varepsilon=(\sqrt{\vphantom{ }H_{12}} - \sqrt{H_{21}}\,)\, / \,
(\sqrt{H_{12}} + \sqrt{H_{21}} \, )$.
    The eigenvalues $\lambda_{1,2} $ and eigenvectors
$|\,\Psi_{1,2}\!> $  of matrix $H$ are
    \begin{equation}
\lambda_{1,2} = H_{11} \pm \frac{H_{12}+H_{21}}{2} \,
\frac{1-\varepsilon^2}{1+\varepsilon^2} \,,
\end{equation}
    \begin{equation}
|\,\Psi_{1,2}\!>=\frac{1}{\sqrt{2\,(1+|\varepsilon |^2)}}\,
\biggl[(1+\varepsilon) \,|\,1\!> \pm \,(1- \varepsilon) \, |\,2\!>\biggr].
\end{equation}
         In particular
\begin{equation}
H=     \left(
\begin{array}{cc}
E-\frac{i}{4}\left(\tau_q^{-1} +\tau_l^{-1}\right)
  &
\frac{1+\varepsilon}{1-\varepsilon}
\left[A-\frac{i}{4}\left(\tau_q^{-1} -\tau_l^{-1}\right)\right]
 \\  & \\
\frac{1-\varepsilon}{1+\varepsilon}
\left[A-\frac{i}{4}\left(\tau_q^{-1} -\tau_l^{-1}\right)\right]
 &
E-\frac{i}{4}\left(\tau_q^{-1} +\tau_l^{-1}\right) \\
\end{array}        \right) .
\label{HM}
\end{equation}

Then the state $|\,\Psi_1 \!> $ describes
short living particles with the life time
$ \ \tau_q \ $ and mass $E+A$.
The state $\ |\,\Psi_2 \!> $ is the state of long living particles
with life time $ \tau_l \ $  and mass $E-A$.
    Here $A$ is the arbitrary parameter $-E<A<E$  and it can be zero,
$E=M_X$.

If $ d = 1 - |\!<\!\Psi_1\,|\,\Psi_2\!>\!|^2 = 1 - |2\, {\rm Re}\,
\varepsilon /(1+|\varepsilon |^2)|^2 $ is the relative part of
long living particles and $\varepsilon $ is real, then $
\varepsilon=(1-\sqrt{d}\,)/\sqrt{1-d}$ \ ($\varepsilon \sim
1-10^{-7} $ for $M_X \sim 10^{14}$\,Gev). So one has a typical
example of "fine tuning" in order to obtain the desired result.

Taking $\tau_q=10^{-35}$ sec and $\tau_l=2\cdot 10^{22}$ sec, one obtains
that for the Hermitean part $(H+H^{+})/2$ of $H$ nondiagonal
$CP$-noninvariant term equal to
$ i\, (\tau_q^{-1} - \tau_l^{-1})\, \varepsilon / (1-\varepsilon^2)/2 $
is of the order of Planckean mass $10^{19}$ Gev.  \\
%
{\it Acknowledgements}.
This work was supported by Min. of Education of Russia, grant
E00-3-163.

%


\begin{thebibliography}{1} \itemsep=-5pt
\bibitem{GMM1}
Grib, A.A., Mamayev, S.G. and Mostepanenko, V.M. 1980,
Quantum Effects in Intensive External Fields,
Atomizdat, Moscow

\bibitem{GMM}
Grib, A.A., Mamayev, S.G. and Mostepanenko, V.M. 1994,
Vacuum Quantum Effects in Strong Fields,
Friedmann Laboratory Publishing, St.Petersburg

\bibitem{Pv}
Pavlov, Yu.V. 2001, Theor. Math. Phys. {\bf 126}, 92

\bibitem{Pv2}
Pavlov, Yu.V. 2001, Theor. Math. Phys. {\bf 128}, 1034

\bibitem{BD}
Birrell, N.D. and Davies, P.C.W. 1982,
Quantum Fields in Curved Space,
Cambridge University Press, Cambridge

\bibitem{ZlSt}
Zel'dovich, Ya.B. and Starobinsky, A.A. 1971, ZhETF {\bf 61}, 2161

\bibitem{GD}
Grib, A.A. and Dorofeev, V.Yu. 1994,  Int. J. Mod. Phys. {\bf D3}, 731

\bibitem{K3}
Kuzmin, V.A. and Tkachev, I.I. 1998, JETP Lett. {\bf 68}, 271

\bibitem{K4}
Kuzmin, V.A. and Tkachev, I.I. 1999,
Ultra High Energy Cosmic Rays and Inflation Relics, hep-ph/9903542

\bibitem{BBV}
Berezinsky, V., Blasi, P. and Vilenkin, A. 1998, Phys. Rev. {\bf D58}, 103515

\end{thebibliography}
\end{document}